# Purely Antiferromagnetic Magnetoelectric Random Access Memory

Tobias Kosub[1,2], Martin Kopte[1,2], Ruben Hühne[3], Patrick Appel[4], Brendan Shields[4], Patrick Maletinsky[4], René Hübner[2], Maciej Oskar Liedke[5], Jürgen Fassbender[2], Oliver G. Schmidt[1], Denys Makarov[1,2]

Magnetic random access memory schemes employing magnetoelectric coupling to write binary information promise outstanding energy efficiency. We propose and demonstrate a purely antiferromagnetic magnetoelectric random access memory (AF-MERAM) that offers a remarkable 50 fold reduction of the writing threshold compared to ferromagnet-based counterparts, is robust against magnetic disturbances and exhibits no ferromagnetic hysteresis losses. Using the magnetoelectric antiferromagnet $Cr_2O_3$, we demonstrate reliable isothermal switching via gate voltage pulses and all-electric readout at room temperature. As no ferromagnetic component is present in the system, the writing magnetic field does not need to be pulsed for readout, allowing permanent magnets to be used. Based on our prototypes of these novel systems, we construct a comprehensive model of the magnetoelectric selection mechanism in thin films of magnetoelectric antiferromagnets. We identify that growth induced effects lead to emergent ferrimagnetism, which is detrimental to the robustness of the storage. After pinpointing lattice misfit as the likely origin, we provide routes to enhance or mitigate this emergent ferrimagnetism as desired. Beyond memory applications, the AF-MERAM concept introduces a general all-electric interface for antiferromagnets and should find wide applicability in purely antiferromagnetic spintronics devices.

[1]Institute for Integrative Nanosciences, Institute for Solid State and Materials Research (IFW Dresden e.V.), 01069 Dresden, Germany. [2]Helmholtz-Zentrum Dresden-Rossendorf e.V., Institute of Ion Beam Physics and Materials Research, 01328 Dresden, Germany. [3]Institute for Metallic Materials, Institute for Solid State and Materials Research (IFW Dresden e.V.), 01069 Dresden. [4]Department of Physics, University of Basel, 4056 Basel, Switzerland. [5]Helmholtz-Zentrum Dresden-Rossendorf e.V., Institute of Radiation Physics, 01328 Dresden, Germany. Correspondence and requests should be addressed to T.K. (email: t.kosub@hzdr.de) and D.M. (email: d.makarov@hzdr.de).

In the effort to develop low power data processing and storage devices, non-volatile random access memory schemes have received considerable attention[1]. Magnetic elements like the magnetic random access memory (MRAM) [**Figure 1**(a)] promise excellent speed, superior rewritability and small footprints, which has led to strong commercial interest in this technology for memory applications. In addition to ferromagnetic MRAM, two complementary approaches have recently emerged for advancing beyond conventional MRAM elements in terms of its writing power and data robustness. On the one hand, switching and reading the antiferromagnetic order parameter of metallic antiferromagnets with charge currents[2,3] has enabled purely antiferromagnetic MRAM (AF-MRAM), granting superior data stability against large magnetic disturbances and potentially even faster switchability. On the other hand, magnetoelectric random access memory (MERAM) promises energy efficient writing of antiferromagnets, by eliminating the need for charge currents through the memory cell and instead relying on electric field induced writing. Reading out the antiferromagnetic state from MERAM has presented a challenge to date as magnetoelectric antiferromagnets (e.g. $BiFeO_3$ or $Cr_2O_3$) are dielectrics. Therefore, the readout signal of MERAM cells is conventionally acquired from a ferromagnet that is coupled to the magnetoelectric antiferromagnet by exchange bias[4–8]. While ferromagnets enable readability, their presence strongly interferes with the magnetoelectric selection of the antiferromagnetic order parameter[9]. This is related to exchange bias and ferromagnetic hysteresis, both of which need to be overcome in the writing process of MERAM with ferromagnets.

Here, we put forth the new concept of purely antiferromagnetic MERAM (AF-MERAM[10]) [**Figure 1**(a)], which avoids the issues associated with the presence of ferromagnets by instead using polarizable paramagnets, e.g. Pt, to probe the order parameter of the magnetoelectric antiferromagnet. As shown schematically in **Figure 1**(b), the prototypical memory cell is comprised of an active layer of insulating magnetoelectric antiferromagnet, a bottom gate electrode for writing purposes, and a top electrode that provides the readout interface via anomalous Hall measurements[11]. Using $Cr_2O_3$ as an AF element, we demonstrate a complete working AF-MERAM cell, proving that this concept yields substantial improvements in terms of magnetoelectric performance over comparable MERAM realizations with ferromagnets. In particular, by removing the ferromagnetic component from MERAM, we reduce the writing threshold by a factor of about 50. These characteristics render AF-MERAM a promising new member to the emerging field of purely antiferromagnetic spintronics[3,12]. We show the magnetoelectric writing and all-electric reading operations of a cell at room temperature over hundreds of read-write cycles. While nonvolatile solid state memory is one possible application of AF-MERAM cells, the concept is applicable to other fields of antiferromagnetic spintronics, such as logics, magnonics[13] and material characterization.

**Results**

**Room temperature operation of AF-MERAM.** To realize the memory cell, we use an epitaxial layer stack of Pt(20 nm)/$\alpha$-$Cr_2O_3$(200 nm)/Pt(2.5 nm) that is prepared on $Al_2O_3$(0001) substrates. Similar stacks with $\alpha$-$Cr_2O_3$ have been extensively studied in the scope of traditional MERAM elements with ferromagnetic Co layers[4,5,14–16]. The thicker bottom Pt film serves as the gate electrode and the thin Pt top layer is used to measure the AF order parameter all-electrically via zero-offset anomalous Hall magnetometry[11] (hereafter zero-offset Hall). This readout approach makes use of the uncompensated boundary magnetization of $\alpha$-$Cr_2O_3$(0001), which is rigidly coupled to the AF bulk and creates proximity magnetization in the Pt film[17,18]. An individual magnetoelectric element is obtained by patterning the top Pt layer. **Figure 1**(c) shows the protocol of an isothermal magnetoelectric switching experiment that was carried out at 19°C in a permanent magnetic field of $H \approx +0.5$ MA/m along the film normal. The test sequence mimics random access operations comprising the three essential elements of any memory cell: writing, storage and reading. One of the key technological advantages is that the memory cell operates in static magnetic fields and writing operations are triggered by the application of a voltage. No energy input is necessary during the storage times. The reproducibility of this process is demonstrated over 300 write-store-read cycles in **Figure 1**(d), during which the cell reveals no performance degradation.

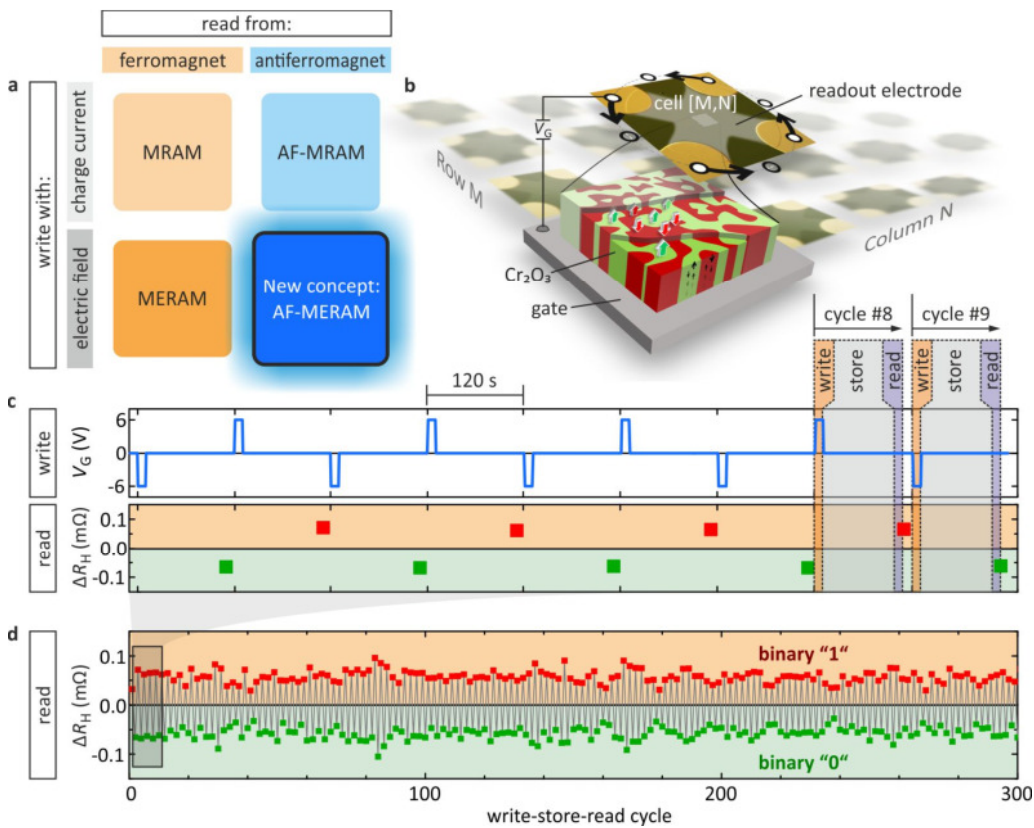

**Figure 1 | Electric field-driven manipulation of the antiferromagnetic order parameter. a**, Non-volatile magnetic random access memory elements categorized according to their writing and readout interfaces. Antiferromagnetic Magnetoelectric Random Access Memory (AF-MERAM) initiates a new field of antiferromagnetic spintronics. **b**, Sketch of one memory cell within a matrix of devices. The arrows indicate the contact permutation to obtain offset free Hall readings[11]. **c**, Random access memory operation where binary information is written by a voltage pulse and stored in the antiferromagnetic order parameter. The magnetic state is read out at a later time after the writing stimulus is removed. **d**, Device behavior over 300 write-store-read cycles.

Two key material requirements must be satisfied to achieve reliable magnetoelectric reversal processes such as shown in **Figure 1**. First, the order parameter has to be susceptible to the gate voltage via the linear magnetoelectric effect. Secondly, the cell has to exhibit thermal stability at the operation temperature, giving rise to stable remanent magnetic states. Both criteria can be directly probed in our system by using the electrical writing and reading interfaces of the magnetoelectric cell. To reveal the exact influence of magnetic and electric field on the antiferromagnetic order parameter, it is mandatory to avoid the influence of magnetic anisotropy, which fixes the order parameter while below the ordering temperature, and instead carry out magnetoelectric field cooling through the ordering temperature. The map in **Figure 2**(a) shows the resulting average antiferromagnetic order parameter in the cell after cooling from 30°C to 7°C using the indicated combination of magnetic cooling field $H_\text{cool}$ and electric cooling field $E_\text{cool} = V_\text{cool}/t$ ($t$ denotes the AF film thickness). For large $EH$ fields, the order parameter selection is consistent with that expected in $\alpha$-$Cr_2O_3$[7,14,19,20] due to the linear magnetoelectric effect. However, for small writing voltages that are technologically desirable, the $EH$ symmetry is disturbed, giving rise to magnetic field-induced selection of the order parameter. Strikingly, the $EH$ symmetry is perfectly restored when accounting for a gate bias voltage $V_\text{GB}$, which is about $-1E$V for this system.

When applying a writing voltage to the cell at 19°C, the antiferromagnetic order parameter can be switched hysteretically with a coercive gate voltage $V_\text{C}$ of about 1.5 V [**Figure 2**(b)], completing the list of ingredients for the non-volatile AF-MERAM prototype. The slightly asymmetric shape of the hysteresis loop is due to the gate voltage range being symmetric about $V_\text{G} = 0$, instead of $V_\text{G} = V_\text{GB}$. The temperature window, in which magnetoelectric writing can be carried out, is limited at higher temperatures by the collapse of antiferromagnetic order and at lower temperatures by magnetic anisotropy[16]. It should be noted that this writability window can be considerably widened to more than 100 K. The high temperature limit can be enhanced by doping[21,22], and the lower temperature limit by applying higher writing voltages[4,5] or by intentionally reducing the anisotropy via doping[21] [SI, section I].

**Table 1** contains an overview of state-of-the-art studies of magnetoelectric functionality using magnetoelectric thin films[4–6,9,14] and single crystals[7], but in both cases relying on interfacial exchange bias with a thin ferromagnetic layer. In addition, the AF-MERAM cell presented in this work is included for comparison. The metrics in the overview are the magnetoelectric film thickness, the writing threshold $(VH)_\text{C}$ and the coercive gate voltage $V_\text{C}$. For integration in microelectronics, the latter two are of particular relevance as the circuit voltage rating depends on them.

| Study | $t$ (μm) | $(VH)_\text{C}$ (MW m$^{-1}$) | $V_\text{C}$ (V) | Magnetic field |
|---|---|---|---|---|
| exchange bias reversal*, Cr$_2$O$_3$/Co/Pd[7] | 1,000 | 240 | 450** | writing pulse |
| exchange bias reversal*, Cr$_2$O$_3$/Co/Pt[5] | 0.2 | 40 | 56** | writing pulse |
| exchange bias reversal*, Cr$_2$O$_3$/Co/Pt[4] | 0.5 | 48 | 105** | writing pulse |
| magnetization switching, BiFeO$_3$/CoFe[6] | 0.1 | --- | 4 | must be ≈0 for readout |
| magnetization switching, Cr$_2$O$_3$/Pt (present work) | 0.2 | 0.75 | 1.5 | permanent |

**Table 1 | Performance chart of MERAM systems.** Overview of state-of-the-art isothermal magnetoelectric switching studies using either the linear magnetoelectric effect in Cr$_2$O$_3$ or the multiferroic coupling in BiFeO$_3$. The value $(VH)_\text{C}$ gives the magnetoelectric writing threshold (product of magnetic field and voltage). The writing voltage $V_\text{C}$ allows to qualitatively compare Cr$_2$O$_3$ based systems and BiFeO$_3$ systems in terms of the voltage at which the magnetization state switches. *Application of the writing voltage does not switch the ferromagnetic Co, but only the antiferromagnetic Cr$_2$O$_3$ implying that the magnetic field must be removed for read-out from the ferromagnet. **For comparability, the writing voltages are calculated for a magnetic field of $H_\text{write} = 0.5$ MA m$^{-1}$ as was used in the present study. The actual used writing voltages in these studies are similar to the normalized values, as the magnetic fields were also similar.

While exchange bias has traditionally been used to probe the antiferromagnetic state of Cr$_2$O$_3$, this leads to strongly increased magnetoelectric coercivities, especially for thin films of Cr$_2$O$_3$[4,5]. When judging the writing threshold, all the exchange bias systems require very large $VH$ for isothermal magnetoelectric switching of the AF order parameter in $Cr_2O_3$. In contrast, the AF-MERAM approach provides a route to reduce both the coercivity and the resulting write voltage by a factor of about 50 over exchange biased examples[SI, section II]. Additionally, AF-MERAM can be read out in permanent external magnetic fields, whereas exchange biased MERAM requires the removal of the magnetic field for readout. Thus, the example here presented opens an appealing field of AF-MERAM with ultra-low writing thresholds and superior stability and readability of the magnetic information in the presence of external magnetic fields.

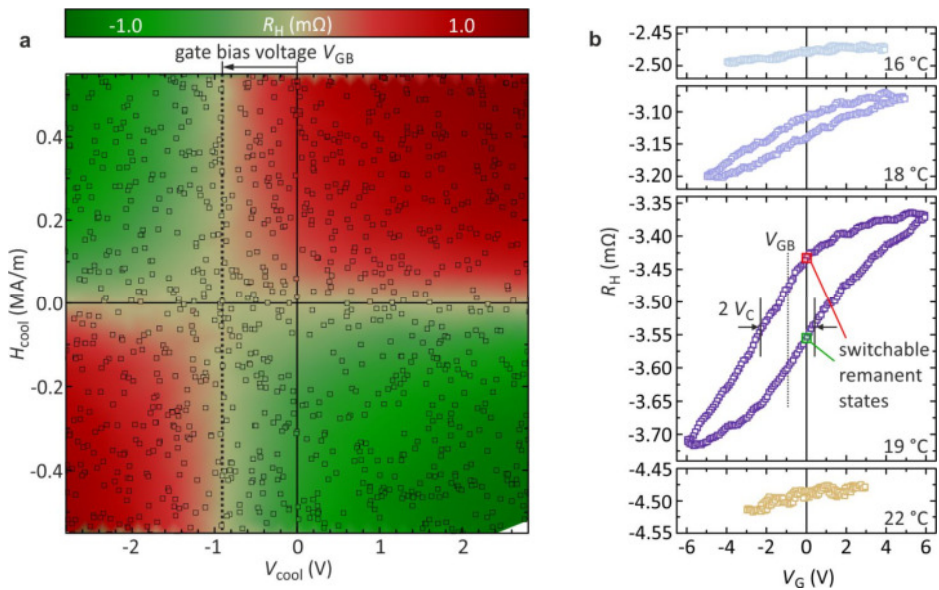

**Figure 2 | Isothermal and field-cooled magnetoelectric selection. a**, Map of the antiferromagnetic state selected by a range of magnetic field and gate voltage combinations during cool-down from 30°C through the antiferromagnetic ordering temperature to the measurement temperature of 7°C. Measurements were carried out at $H = 0$ and $V_G = 0$. The squares are data points and the background color is a linear interpolation. **b**, Readout signal corresponding to the antiferromagnetic order parameter of the cell as a function of the writing voltage $V_G$ for several temperatures near the antiferromagnetic ordering temperature and $H = 0.5\,\mathrm{MA\,m^{-1}}$. The open hysteresis loop with coercivity $V_C$ gives rise to switchable remanent states.

## Discussion

**Gate bias voltage.** The gate bias voltage of $V_{\text{GB}} \approx -1\text{V}$ [**Figure 2**] presents a key challenge for achieving ultra-low voltage threshold switching and ultra-high data stability. It has a detrimental effect on both the required writing voltage and on the data stability at zero voltage as the antiferromagnetic state develops a susceptibility to magnetic fields, even in the absence of an electric field ($V_{\text{G}} = 0$). We find that the gate bias voltage is a material characteristic in thin films of magnetoelectric antiferromagnets, and in the following we reveal its physical origin and derive a means to control its value.

When combining the large body of data on $Cr_2O_3$ thin film systems[4,5,7,9,14,16,18,23–30], a coherent picture emerges: the total magnetoelectric energy density exerting a selection pressure on the antiferromagnetic order parameter in thin film magnetoelectric antiferromagnets is composed of three effects that act simultaneously:

$$U_{\text{MEAF}}(E, H) = t^{-1}\alpha V_{\text{G}} H + t^{-1} J_{\text{EB}} + t^{-1} \rho_m \mu_0 H \tag{1}$$

The first term describes the linear magnetoelectric effect with its coefficient $\alpha$ reported to be approximately $1\text{ ps m}^{-1}$ in $Cr_2O_3$[19,31]. This is the only desired effect in the context of MERAM devices, while the other two effects are parasitic. The second term is the influence of the exchange bias coupling strength $J_{\text{EB}}$ on the antiferromagnet. While this term was typically the strongest contribution in previous studies [SI, section III], it is zero in AF-MERAM due to the lack of a ferromagnet. The last term arises from a non-zero areal magnetic moment density $\rho_m$ within the antiferromagnet itself, which renders the material ferrimagnetic. This term, due to emergent ferrimagnetism, cannot be excluded a priori. The gate bias voltage can now be calculated from Eq. (1):

$$V_{\text{GB}} = -\mu_0 \frac{\rho_m}{\alpha} \tag{2}$$

The gate bias voltage $V_{GB}$ is in an intimate relation with the magnetoelectric coefficient $\alpha$ and the areal magnetic moment density $\rho_m$ at the onset temperature of the thermal stability of the antiferromagnetic order. Eq. (2) implies that the gate bias in magnetoelectric field cooling experiments vanishes for perfectly antiferromagnetic order ($\rho_m = 0$). Its non-zero value in our system can be used to estimate the approximate ferrimagnetic moment density at the ordering temperature of about 21°C, yielding a value of $\rho_m \approx 0.1$ $\mu_B$ nm$^{-2}$. Conversely, achieving a low gate bias voltage requires that ferrimagnetism is averted.

**Origin and tunability of emergent ferrimagnetism.** The presence of ferrimagnetism cannot be accounted for by any intrinsic effect within the Cr$_2$O$_3$ antiferromagnetic film, as all magnetic moments, including boundary moments[11,17,18], are intrinsically compensated when accounting for all boundaries [SI, section IV, **Supplementary Figure 1**]. Therefore, extrinsic effects are necessary to break the sublattice equivalence and produce ferrimagnetism. In the following, we present an in-depth study of extrinsic thin film phenomena and their influence on the emergent ferrimagnetism. Namely, we invoke different degrees of crystalline twinning, elastic lattice deformation, intermixing and misfit dislocation density in Cr$_2$O$_3$ thin films, by preparing three distinct systems with epitaxial underlayers of Al$_2$O$_3$(0001), Pt(111) or V$_2$O$_3$(0001).

One striking result is that the gate bias voltage, and thus the emergent ferrimagnetism, can indeed be controlled by the choice of underlayer material. In particular, when Cr$_2$O$_3$ thin films are grown on a V$_2$O$_3$ underlayer, ferrimagnetism is almost entirely eliminated. **Figure 3**(a) shows a magnetoelectric field cooling map of the V$_2$O$_3$ buffered system exhibiting virtually perfect $EH$ symmetry. As highlighted by the indicated line profiles [**Figure 3**(b,c)], the selection preference for a particular antiferromagnetic state vanishes when either $E = 0$ or $H = 0$, as expected from the pristine action of the linear magnetoelectric effect [first term in Eq. (1)]. The possibility to completely eliminate the gate bias is highly relevant for AF-MERAM applications, as the AF state can then be switched with lower voltages [**Supplementary Figure 2**] and is completely stable once the gate voltage returns to zero.

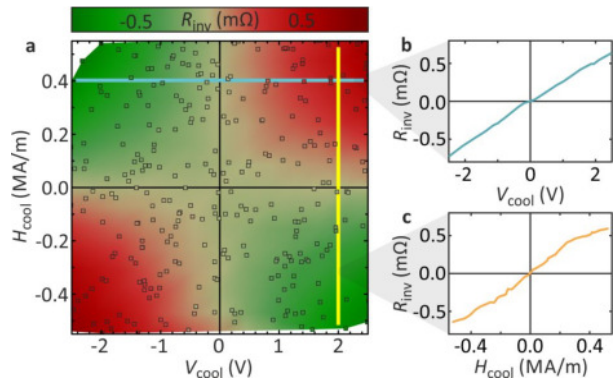

**Figure 3 | Influence of a $V_2O_3$ buffer layer on $Cr_2O_3$ magnetoelectricity. a**, Magnetoelectric field cooling map of the antiferromagnetic order selection for a $V_2O_3/Cr_2O_3/Pt$ system. **b** and **c**, Line profiles taken from the map in panel a. Only non-zero products of gate voltage and magnetic field lead to appreciable order parameter selection, whereas the individual stimuli do not.

To pinpoint the specific extrinsic effect that is responsible for the different degrees of emergent ferrimagnetism in $Cr_2O_3$ thin films grown on different underlayers, it is instructive to correlate the observed areal magnetic moment density and the various growth-induced effects [**Table 2**]. The areal magnetization is determined via the slope of the dependence of the antiferromagnetic order parameter selection on the magnetic field [**Figure 4**(c), **Supplementary Figure 5**]. The gradual shape of these dependences results from a selection tendency of uniaxial antiferromagnetic domains according to their ferrimagnetism, averaged over the readout electrode area. To verify that the microscopic ordering is indeed a mixture of purely uniaxial domains, images of the surface magnetization states after zero-field cooling [**Figure 4**(d)] and field-cooling [**Figure 4**(b)] were obtained by scanning nitrogen-vacancy magnetometry[32–34]. This technique measures the stray magnetic field $\approx 50$ nm above the sample surface, clearly indicating the equal presence of up- and down-domains in the zero-field cooled state. In contrast, field-cooling predominantly selects one of the two domain orientations which allows to calculate the degree of ferrimagnetism in the films.

It should also be noted, that measuring the gate bias voltage $V_{GB}$ [Eq. (2)] provides a second route to quantify the areal magnetization $\rho_m$ absolutely, which is however restricted to conducting underlayers and suffers from the uncertainty of the value of $\alpha$. Therefore, magnetization values determined via the gate bias will not be used for the comparison of the different underlayer materials.

| Underlayer material | Twinning ratio | $c$-axis compression | Expected miscibility | Linear misfit | Areal magnetization $\rho_m$ (a.u.) |
|---|---|---|---|---|---|
| $Al_2O_3$ | ≈ 2 % | ≈ 0.0 % | weak | +4.0 % | +0.455 ± 0.28 |
| Pt | ≈ 50 % | 0.18 % | none | +2.8 % | +0.100 ± 0.043 |
| $V_2O_3$ | ≈ 2 % | 0.30 % | strong | −0.5 % | −0.0021 ± 0.001 |

**Table 2 | Influence of different underlayers on structural and ferrimagnetic properties of $Cr_2O_3$ thin films.** The values for the structural properties are derived in detail in the [SI, section VI, **Supplementary Figure 3** and **Supplementary Figure 4**]. The ferrimagnetic moment density values are relative values obtained by zero-offset Hall [SI, sections VII, **Supplementary Figure 5** and **Supplementary Figure 6**]. They are normalized to the approximate value for $Pt/Cr_2O_3/Pt$ obtained via the gate bias voltage as of Eq. (2).

Based on these data, we conclude that elastic film strain, twinning and cation intermixing in epitaxial $Cr_2O_3$ films cannot account for the observed degree of ferrimagnetism, as none of these properties are correlated to the areal magnetization [**Table 2**]. Instead, the results suggest that the lattice mismatch is the cause of the emergent ferrimagnetism.

The scaling relationship between the measured areal magnetic moment density and the linear lattice misfit between $Cr_2O_3$ and its underlayer is shown in **Figure 4**(a). When taking into account the data of the three investigated systems, we find that the data align tightly to a quadratic scaling relation (red line). Such a relationship hints at the number of the misfit dislocations per area being the key property determining the areal ferrimagnetic moment density. This result leads to a picture in which the population of the two antiferromagnetic sublattices is unbalanced by the presence of misfit dislocations.

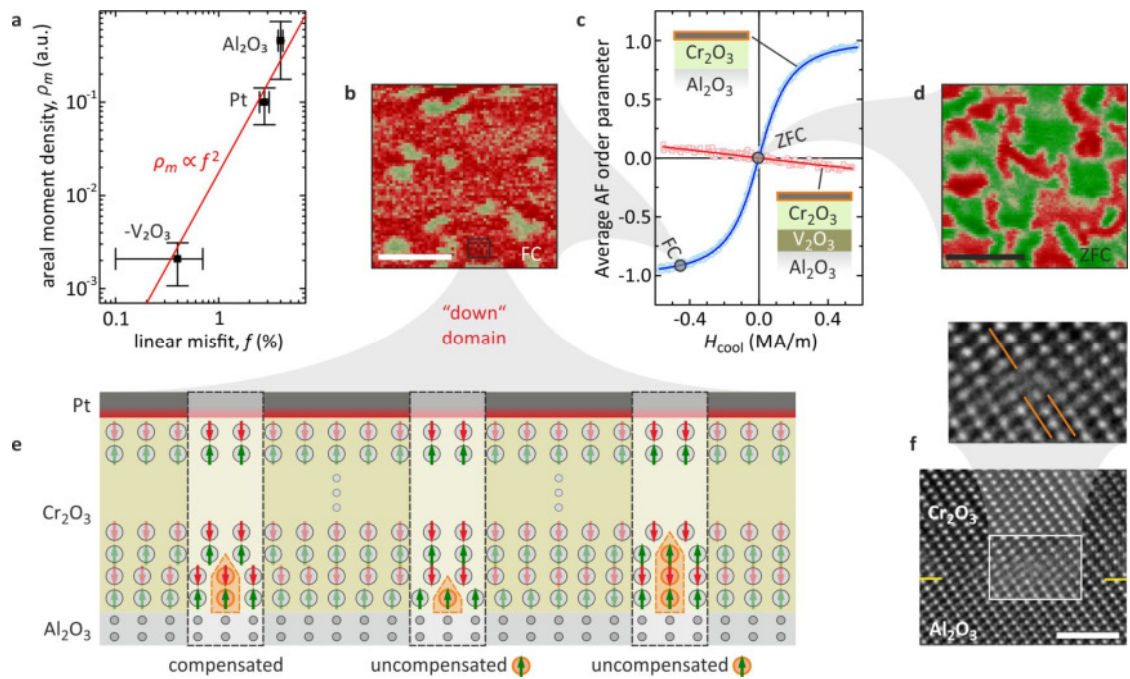

**Figure 4 | Thin film $Cr_2O_3$ behaves ferrimagnetically. a**, Scaling between the measured areal magnetic moment density and the linear misfit between $Cr_2O_3$ and its underlayer. **b** and **d**, Images of the surface magnetic stray field after field-cooling (FC) and zero-field-cooling (ZFC), respectively, were obtained by scanning nitrogen vacancy microscopy (see text). Scale bars are 1 µm. **c**, The emergent ferrimagnetism couples strongly to external magnetic fields and renders the antiferromagnetic order parameter selectable by magnetic fields much smaller than anisotropy fields. **e**, Sketch of the effect of misfit dislocations on the atomic population of the two antiferromagnetic sublattices. **f**, HRTEM images of the $Al_2O_3/Cr_2O_3$ interface (yellow guide lines) showing complete structural coherence disrupted by occasional misfit dislocations. The scale bar is 2 nm.

Such dislocations arise due to the heteroepitaxial film growth [**Figure 4**(f)] as the dominant defect type of the otherwise highly coherent interface and appear within the first atomic layers of the $Cr_2O_3$ film as evidenced by positron annihilation spectroscopy [**Supplementary Figure 7**, **Supplementary Table 1**]. As sketched for the case of compressive misfit in **Figure 4**(e), the dislocations (orange boxes) can contain unequal populations of spin "up" and spin "down" atoms if the dislocation terminates after an odd number of atomic layers. These surplus spins are all aligned within one domain due to the layered sublattice structure in $\alpha$-$Cr_2O_3$(0001). While the magnetic moment of atoms near dislocations might be different from atoms in the relaxed lattice, this picture serves to illustrate that misfit dislocations do indeed unbalance the atomic populations in each of the two sublattices.

Remarkably, the lattice misfit not only correlates with the magnitude of the emergent magnetization, but also with its sign with respect to the antiferromagnetic order parameter. This sign change of the ferrimagnetic behavior in the case of tensile misfit for the $V_2O_3$ buffered sample emerges naturally from the previously introduced picture. Tensile misfit results in atoms being skipped from the bottom boundary sublattice instead of atoms being added. Therefore, tensile misfit results in the top boundary magnetization being aligned along the cooling field, while compressive misfit results in the top boundary magnetization being aligned opposite to the magnetic cooling field.

To quantify ferrimagnetism, we investigate the magnetic field-induced antiferromagnetic order parameter selection with no electric field applied [**Figure 4**(c)], which is influenced exclusively by the last term of Eq. (1). The $Al_2O_3$ buffered films are clearly more susceptible to the magnetic field than the $V_2O_3$ buffered films, which is in line with the substantially larger lattice misfit of the former over the latter. Moreover, a magnetic field of the same sign selects opposite antiferromagnetic states in the two systems, which corresponds to the opposite sign of the lattice misfit.

In conclusion, we demonstrated reliable room temperature magnetoelectric random access memory cells based on a new scheme that relies purely on antiferromagnetic components and does not require a ferromagnet for readout. This AF-MERAM provides substantially reduced writing thresholds over conventional MERAM prototypes, enabling further improvements in the energy efficiency of non-volatile solid state memory and logics. Since a permanent magnetic writing field does not interfere with readout in AF-MERAM, this new approach extends voltage driven writing to magnetoelectric antiferromagnets like $Cr_2O_3$, whereas such functionality has previously been feasible only in multiferroic antiferromagnets such as $BiFeO_3$ [**Table 1**]. It should be noted that the advantages of omitting the ferromagnet from MERAM cells likewise apply to multiferroic antiferromagnets, opening an appealing field of AF-MERAM with ultra-low writing thresholds and superior stability of the magnetic order parameter. The concept also provides an important new building block for the emerging field of antiferromagnetic spintronics. While we did not investigate the speed of the actual writing process, first prototypes of conventional MERAM could be switched within a few tens of ns[5].

We use thin films of magnetoelectric antiferromagnetic $Cr_2O_3$ as the core material and find that this material becomes ferrimagnetic when grown as epitaxial thin films. Emergent ferrimagnetism in thin films of magnetoelectric antiferromagnets can be desirable[35]. For the application to purely antiferromagnetic magnetoelectric elements, however, ferrimagnetism should be minimized. Through an in-depth structural characterization, we find that the observed degree of ferrimagnetism is correlated with the square of the linear lattice misfit between $Cr_2O_3$ and its underlayer. This finding provides both a fundamental mechanism for the phenomenon of emergent ferrimagnetism and suggests a readily available tuning knob to enhance or eliminate the magnetic field coupling of magnetoelectric antiferromagnets.

**Methods**

Oxide films were grown by reactive evaporation of the base metal in high vacuum onto *c*-cut sapphire substrates (Crystec GmbH) heated to 700°C initially and to 500°C after the first few monolayers. The background gas used was molecular oxygen at a partial pressure of $10^{-5}$ mbar. Chromium was evaporated from a Knudsen cell, vanadium was evaporated from a block target using an electron-beam and platinum was sputtered from a DC magnetron source. Deposition of the oxides was carried out using rates of about $0.4\,\text{Å}\,\text{s}^{-1}$ and was monitored in situ by reflection high energy electron diffraction (RHEED). Oxide layers were subjected to a vacuum annealing process at 750°C and residual pressure of $10^{-7}$ mbar directly after growth. The thin Pt top layers were deposited at lower temperatures of about 100°C using a higher rate of $1.0\,\text{Å}\,\text{s}^{-1}$ to maintain layer continuity. Hall crosses were patterned from the top Pt layers, by $SF_6$ reactive ion etching around a photoresist mask. Transport was measured using zero-offset Hall[11]. Typical current amplitudes were on the order of $500\,\mu A$. RAM operation was carried out in a permanent magnetic background field of $H \approx +0.5\,\text{MA}\,\text{m}^{-1}$ along the film normal.

To obtain the average AF order parameter dependence on the magnetic cooling field, the data of the spontaneous Hall signal after cooling under a range of field values were fitted by an expression that provides the normalization and the absolute magnetic moment distribution of individual domain pieces within the $Cr_2O_3$ films [SI, section VII]. The relative domain sizes of films with different buffer layers were determined using zero-offset Hall by evaluating the statistics of the domain selection within the finite-size Hall crosses [SI, section VII].

The structural properties of the $Cr_2O_3$ films on different buffer layers were characterized by x-ray diffraction and channeling contrast scanning electron microscopy as shown in detail in [SI, section VI].

Scanning nitrogen-vacancy (NV) magnetometry was performed with a tip fabricated from single-crystal, <100>-oriented diamond that was implanted with $^{14}$N ions at 6 keV, and annealed at 800°C to form NV centers[36]. An external field of $2.2\,\text{kA}\,\text{m}^{-1}$ was applied along the NV axis (diamond <111> crystal direction) to induce Zeeman splitting of the NV electronic ground-state spin. A microwave driving field was then locked to the spin transition at $\approx 2.864\,\text{GHz}$ to track the additional Zeeman shift due to the stray field of the magnetic film surface[37]. Magnetic field values for each pixel were obtained by averaging the microwave lock frequency for 7 s.

**Acknowledgements**

We acknowledge the support from Dr. S. Harazim for the maintenance of the clean room facilities, S. Nestler for conducting the reactive ion etching, M. Bauer for depositing the contact pads, R. Engelhard for the maintenance of the deposition tools and D. Karnaushenko and D. D. Karnaushenko for providing polyimide photoresist (all IFW Dresden). Support by the Structural Characterization Facilities at IBC of the HZDR is gratefully acknowledged. This work was funded in part by the European Research Council under the European Union's Seventh Framework Programme (FP7/2007-2013)/ERC grant agreement n° 306277 and the European Union Future and Emerging Technologies Programme (FET-Open Grant No. 618083).

**Author Contributions**

T.K. prepared the samples. T.K. and M.K. set up the magneto-transport measurement system. T.K. carried out the electrical measurements and the corresponding data analysis. T.K., M.K. and R. Hühne conducted the x-ray diffraction studies. P.A., B.S. and P.M. carried out the nitrogen-vacancy microscopy measurements. R. Hübner imaged cross-sections of the samples in TEM. M.O.L. conducted the PAS investigations. T.K. created the graphics and T.K. and D.M. wrote the manuscript with comments from all authors. D.M., O.G.S. and J.F. supervised the project.

**Competing Financial Interests**

We declare no competing financial interests.

**References**

1. Yang, J. J., Strukov, D. B. & Stewart, D. R. Memristive devices for computing. *Nature Nanotechnology* **8**, 13–24 (2013).
2. Jungwirth, T., Marti, X., Wadley, P. & Wunderlich, J. Antiferromagnetic spintronics. *Nat Nanotechnoly* **11**, 231–241 (2016).
3. Wadley, P. *et al.* Electrical switching of an antiferromagnet. *Science* **351**, 587–590 (2016).
4. Ashida, T. *et al.* Isothermal electric switching of magnetization in $Cr_2O_3$/Co thin film system. *Appl. Phys. Lett.* **106**, 132407 (2015).
5. Toyoki, K. *et al.* Magnetoelectric switching of perpendicular exchange bias in Pt/Co/$Cr_2O_3$/Pt stacked films. *Appl. Phys. Lett.* **106**, 162404 (2015).
6. Heron, J. *et al.* Deterministic switching of ferromagnetism at room temperature using an electric field. *Nature* **516**, 370–373 (2014).
7. He, X. *et al.* Robust isothermal electric control of exchange bias at room temperature. *Nature Materials* **9**, 579–585 (2010).
8. Matsukura, F., Tokura, Y. & Ohno, H. Control of magnetism by electric fields. *Nature Nanotechnology* **10**, 209–220 (2015).
9. Toyoki, K. *et al.* Switching of perpendicular exchange bias in Pt/Co/Pt/$Cr_2O_3$/Pt layered structure using magneto-electric effect. *J. Appl. Phys.* **117**, 17D902 (2015).
10. Kosub, T., Schmidt, O. G. & Makarov, D. Magnetoelektrische Funktionselemente, Patent Applied For. (2015).
11. Kosub, T., Kopte, M., Radu, F., Schmidt, O. G. & Makarov, D. All-electric access to the magnetic-field-invariant magnetization of antiferromagnets. *Phys. Rev. Lett.* **115**, 097201 (2015).
12. Marti, X. *et al.* Room-temperature antiferromagnetic memory resistor. *Nature Materials* **13**, 367–374 (2014).
13. Rovillain, P. *et al.* Electric-field control of spin waves at room temperature in multiferroic $BiFeO_3$. *Nat. Mater.* **9**, 975–979 (2010).
14. Ashida, T. *et al.* Observation of magnetoelectric effect in $Cr_2O_3$/Pt/Co thin film system. *Appl. Phys. Lett.* **104**, 152409 (2014).
15. Iwata, N., Kuroda, T. & Yamamoto, H. Mechanism of Growth of $Cr_2O_3$ Thin Films on (1$\underline{1}$02), (11$\underline{2}$0) and (0001) Surfaces of Sapphire Substrates by Direct Current Radio Frequency Magnetron Sputtering. *Jap. J. Appl. Phys.* **51**, 11PG12 (2012).
16. Fallarino, L., Berger, A. & Binek, C. Magnetic field induced switching of the antiferromagnetic order parameter in thin films of magnetoelectric chromia. *Phys. Rev. B* **91**, 054414 (2015).
17. Belashchenko, K. D. Equilibrium Magnetization at the Boundary of a Magnetoelectric Antiferromagnet. *Phys. Rev. Lett.* **105**, 147204 (2010).
18. Wu, N. *et al.* Imaging and Control of Surface Magnetization Domains in a Magnetoelectric Antiferromagnet. *Phys. Rev. Lett.* **106**, 087202 (2011).
19. Fiebig, M. Revival of the magnetoelectric effect. *J. Phys. D: Appl. Phys.* **38**, R123 (2005).
20. Dzyaloshinskii, I. E. On the magneto-electrical effect in antiferromagents. *Sov. Phys.–JETP* **37**, 881–882 (1959).
21. Mu, S., Wysocki, A. L. & Belashchenko, K. D. Effect of substitutional doping on the Néel temperature of $Cr_2O_3$. *Phys. Rev. B* **87**, 054435 (2013).
22. Street, M. *et al.* Increasing the Néel temperature of magnetoelectric chromia for voltage-controlled spintronics. *Appl. Phys. Lett.* **104**, 222402 (2014).
23. Shiratsuchi, Y., Fujita, T., Oikawa, H., Noutomi, H. & Nakatani, R. High Perpendicular Exchange Bias with a Unique Temperature Dependence in Pt/Co/α-$Cr_2O_3$(0001) Thin Films. *Appl. Phys. Exp.* **3**, 113001 (2010).

24. Shiratsuchi, Y. *et al.* High-Temperature Regeneration of Perpendicular Exchange Bias in a Pt/Co/Pt/α-$Cr_2O_3$/Pt Thin Film System. *Appl. Phys. Exp.* **6**, 123004 (2013).
25. Fallarino, L., Berger, A. & Binek, C. Giant temperature dependence of the spin reversal field in magnetoelectric chromia. *Appl. Phys. Lett.* **104**, 022403 (2014).
26. Nozaki, T. *et al.* Positive exchange bias observed in Pt-inserted $Cr_2O_3$/Co exchange coupled bilayers. *Appl. Phys. Lett.* **105**, 212406 (2014).
27. Shiratsuchi, Y., Nakatani, T., Kawahara, S. & Nakatani, R. Magnetic coupling at interface of ultrathin Co film and antiferromagnetic $Cr_2O_3$(0001) film. *J. Appl. Phys.* **106**, 033903 (2009).
28. Lim, S.-H. *et al.* Exchange bias in thin-film $(Co/Pt)_3$/$Cr_2O_3$ multilayers. *J. Magn. Magn. Mater.* **321**, 1955–1958 (2009).
29. Sahoo, S. & Binek, C. Piezomagnetism in epitaxial $Cr_2O_3$ thin films and spintronic applications. *Phil. Mag. Lett.* **87**, 259–268 (2007).
30. Shiratsuchi, Y. *et al.* Detection and In Situ Switching of Unreversed Interfacial Antiferromagnetic Spins in a Perpendicular-Exchange-Biased System. *Phys. Rev. Lett.* **109**, 077202 (2012).
31. Folen, V. J., Rado, G. T. & Stalder, E. W. Anisotropy of the Magnetoelectric Effect in $Cr_2O_3$. *Phys. Rev. Lett.* **6**, 607–608 (1961).
32. Balasubramanian, G. *et al.* Nanoscale imaging magnetometry with diamond spins under ambient conditions. *Nature* **455**, 648–651 (2008).
33. Maletinsky, P. *et al.* A robust scanning diamond sensor for nanoscale imaging with single nitrogen-vacancy centres. *Nature Nanotechnology* **7**, 320–324 (2012).
34. Taylor, J. *et al.* High-sensitivity diamond magnetometer with nanoscale resolution. *Nature Physics* **4**, 810–816 (2008).
35. Halley, D. *et al.* Size-induced enhanced magnetoelectric effect and multiferroicity in chromium oxide nanoclusters. *Nat. Commun.* **5**, 3167 (2014).
36. Appel, P. *et al.* Fabrication of all diamond scanning probes for nanoscale magnetometry. *arXiv:1604.00021* (2016).
37. Schoenfeld, R. S. & Harneit, W. Real Time Magnetic Field Sensing and Imaging Using a Single Spin in Diamond. *Phys. Rev. Lett.* **106**, 030802 (2011).
38. Martin, T. & Anderson, J. Antiferromagnetic domain switching in $Cr_2O_3$. *IEEE Trans. Magn.* **2**, 446–449 (1966).
39. *Single Crystal Sapphire*. (Kyocera Corp. Fine Ceramics Group: 2014).at <http://global.kyocera.com/prdct/fc/product/pdf/s_c_sapphire.pdf>
40. Edsinger, R. E., Reilly, M. L. & Schooley, J. F. Thermal Expansion of Platinum and Platinum–Rhodium Alloys. *J. Res. Natl. Bur. Stand.* **91**, 333–356 (1986).
41. Eckert, L. J. & Bradt, R. C. Thermal expansion of corundum structure $Ti_2O_3$ and $V_2O_3$. *J. Appl. Phys.* **44**, 3470–3472 (1973).
42. Zhang, L., Kuhn, M. & Diebold, U. Epitaxial growth of ultrathin films of chromium and its oxides on Pt (111). *J. Vac. Sci. Tech. A* **15**, 1576–1580 (1997).
43. Kim, S. S. & Sanders, T. H. Thermodynamic Modeling of the Isomorphous Phase Diagrams in the $Al_2O_3$–$Cr_2O_3$ and $V_2O_3$–$Cr_2O_3$ Systems. *J. Am. Ceram. Soc.* **84**, 1881–1884 (2001).
44. Bayati, M. *et al.* Domain epitaxy in $TiO_2$/α-$Al_2O_3$ thin film heterostructures with $Ti_2O_3$ transient layer. *Appl. Phys. Lett.* **100**, 251606 (2012).
45. Kaneko, K., Kakeya, I., Komori, S. & Fujita, S. Band gap and function engineering for novel functional alloy semiconductors: Bloomed as magnetic properties at room temperature with α-$(GaFe)_2O_3$. *J. Appl. Phys.* **113**, 233901 (2013).
46. Dehm, G., Inkson, B. & Wagner, T. Growth and microstructural stability of epitaxial Al films on (0001) α-$Al_2O_3$ substrates. *Acta Mater.* **50**, 5021–5032 (2002).
47. Liedke, M. O. *et al.* Open volume defects and magnetic phase transition in Fe60Al40 transition metal aluminide. *J. Appl. Phys.* **117**, 163908 (2015).

48. Saleh, A. S. Analysis of positron profiling data by ROYPROF, VEPFIT, and POSTRAP4 codes: a comparative study. *J. Theo. Appl. Phys.* **7**, 1–6 (2013).
49. Van Veen, A., Schut, H., De Vries, J., Hakvoort, R. & Ijpma, M. Analysis of positron profiling data by means of 'VEPFIT. *4th International workshop on: Slow-positron beam techniques for solids and surfaces* **218**, 171–198 (1991).
50. Gordo, P. M. *et al.* On the defect pattern evolution in sapphire irradiated by swift ions in a broad fluence range. *Applied Surface Science* **255**, 254–256 (2008).

## Supplemental Information

### I. Effects of substitutional doping on the magnetic properties of $Cr_2O_3$

The Néel temperature of pristine $Cr_2O_3$ of about 307 K is too low to be worthwhile for commercial room-temperature applications. This issue has previously been investigated in the literature both theoretically[21] and the predictions were later confirmed in experiment qualitatively and quantitatively[22]. Namely, the Néel temperature can be enhanced to roughly 400 K by about 3 % substitutional anion doping of boron for oxygen .Therefore, we consider the predictions regarding intentional cation doping obtained by the same calculations as highly relevant.

As shown in Ref.[21] different dopants display differences in the details of their substitutional effect. Some dopants (Ni, Co) affect nearest neighbor exchange more severely than long range exchange. Some dopants such as Mn or Fe are predicted to increase the average sublattice magnetization while the other dopants are predicted to reduce it. One should also notice, that $\alpha$-$Fe_2O_3$ and $\alpha$-$Ti_2O_3$ are themselves corundum structure antiferromagnets, but with different antiferromagnetic order. In particular, $Fe_2O_3$ possesses a Néel temperature of 950 K, but no linear magnetoelectric effect in its pristine form.

All of these facts lead us to the conclusion, that the substitutional doping of $Cr_2O_3$ is a complex endeavor that can yield various effects like the demonstrated increase of the Néel temperature or the change of the antiferromagnetic order type. Combining different dopants is therefore likely to address both the Néel temperature and the magnetic anisotropy, possibly to different or even inverse extents.

### II. Calculation of the writing threshold reduction factor of AF-MERAM compared to MERAM

Although comparisons between results obtained for thin film system prepared by different groups in different chambers and different fabrications habits are often complicated, the situation is less ambiguous in the case of $Cr_2O_3$-based thin film MERAM prototypes. All of the successful demonstrations of these prototypes are based on the same material system, which yields consistent results throughout several groups. This system is based on (0001) cut Sapphire single crystal substrates, a roughly 20 nm thick sputtered Pt gate layer, a roughly 200 nm thick $Cr_2O_3$ layer prepared at about 600°C and a metallic sensing layer sputtered at room temperature. Throughout the last decade, this last sensing layer has been very thoroughly characterized[4,5,7,9,14,23,24,26–28,30]. One of the key conclusions is that the stronger the exchange bias between $Cr_2O_3$ and a ferromagnetic Co sensing layer, the easier it is to read out the system (due to the larger shift of the ferromagnetic hysteresis loop) but the harder it is to write the antiferromagnet via the magnetoelectric effect[9]. By finely tuning the exchange bias strength via intentional decoupling of $Cr_2O_3$ and Co, several groups eventually managed to achieve both writability of $Cr_2O_3$ and stable exchange bias at low temperatures. The few works which demonstrate this functionality[4,5,7] are all included as performance references in our manuscript and are best-case scenarios selected from a much wider body of research work. Many articles published within the last decade, but also internal work done by us, showed reliable exchange bias between $Cr_2O_3$ and Co, reliable isolation of the gate electrode but no writability by the magnetoelectric effect because the necessary writing voltages were beyond the dielectric breakdown strength of $Cr_2O_3$[28].

On the other hand, magnetoelectric switching of pristine $Cr_2O_3$ without an attached ferromagnet has been performed since the discovery of the linear magnetoelectric effect in the early 1960ies. Many of these experiments yielded even lower values of the $H$ product than reported in the current manuscript as necessary to switch the AF order parameter in crystal samples[38] or to field cool to single domain states in thin film sample[18].

Therefore, it is very likely that by selecting the best-case writing thresholds for conventional MERAM and comparing these figures to the intermediate plain $Cr_2O_3$ writing threshold obtained by our AF-MERAM prototype, we are in fact underestimating the reduction factor of the writing thresholds that is achieved by removing the ferromagnetic layer from a thin film MERAM system. For structurally entirely comparable MERAM and AF-MERAM systems, the reduction factor of the writing threshold could thus be even higher than 50-fold compared to the ferromagnet-containing counterparts.

A vague estimation of the reduction factor that could be achieved by removing the ferromagnet from optimized traditional MERAM system, can be learnt when comparing the energy contributions to the domain selection in magnetoelectric antiferromagnets [SI, section III]. These latter consideration yield an estimated reduction factor of the writing threshold of AF-MERAM compared to traditional MERAM on the order of 1000.

### III. Energy contributions to antiferromagnetic domain selection in magnetoelectric antiferromagnets

To assess the relative importance of the three contributions, exemplary values for the film thickness, the electric and magnetic fields of $t = 100$ nm, $V_G = 1$ V and $H = 0.1$ MA/m will be used. These particular values represent moderate quantities that can be routinely achieved in technological applications.

The selection pressure due to the linear magnetoelectric effect [first term in Eq. (1)] is on the order of 1 Pa for the magnetoelectric coefficient of $Cr_2O_3$ of about $\alpha \approx 1\,\text{ps/m}$[19,31]. The second term describes the interfacial exchange bias coupling energy which depends on the relative alignment between the ferromagnetic magnetization and the AF order parameter. In existing works, the coupling was usually collinear along the film normal. The coupling constant[9] is about $J_{EB} = 0.1\,\text{mJ/m}^2$. This effect generates of a selection pressure of about 1000 Pa for the assumed parameters. The last term is caused by the Zeeman energy which arises as a consequence of non-zero areal magnetic moment density $\rho_m$ in thin films of magnetoelectric antiferromagnet. Such magnetization was attributed to surface effects and was found to be on the order of $\rho_m = 0.1\,\mu_B/\text{nm}^2$ for similar systems[16]. This value can be extracted when assuming a signal level of $10^{-8}$ emu and a sample area of 0.1 cm² as are typical for the device used in Ref. [16]. Calculating the areal magnetization via the gate bias voltage as done in the main text, yields a similar magnitude. With the aforementioned parameters this contribution is also on the order of 1 Pa.

As a result, the exchange bias contribution far outweighs the other contributions at a magnitude of about 1000 Pa. Consequently, the writability of information via the linear magnetoelectric effect [first contribution in Eq. (1)] is severely compromised if exchange bias coupling is present.

## IV.    Hall measurements of both surfaces of the $Cr_2O_3$ thin film

**Supplementary Figure 1** shows magnetic hysteresis loops obtained by zero-offset Hall[11] of $Cr_2O_3$/Pt top and bottom interfaces, respectively. The high temperature paramagnetic curves show a negative slope with the magnetic field implying that the sign of the anomalous Hall signal in Pt is opposite to the sign of the boundary magnetization in $Cr_2O_3$. The reason for that is that either the anomalous Hall coefficient of Pt or the proximity magnetization in Pt with respect to Cr is negative. Magnetic field cooling in a positive field of $H_{cool} = 0.5$ MA/m down to 9°C is then used to induce an almost saturated AF state in the $Cr_2O_3$ film exploiting the weak ferrimagnetism. Such a domain state is expected to display ferromagnetic surface terminations with different signs of the magnetization at the two interfaces. As a result, the observed Hall signals of the two boundaries show opposite sign and almost identical magnitudes when considering the different thicknesses of the Pt layer. When expressed in terms of the Hall resistivities $\rho_H = R_H t_{Pt}$, one obtains $+2.5$ pΩm for the top surface and $-3$ pΩm for the bottom surface. The two readings can be judged to be of similar magnitude, but as two different samples are measured, a quantitative comparison would be ambiguous.

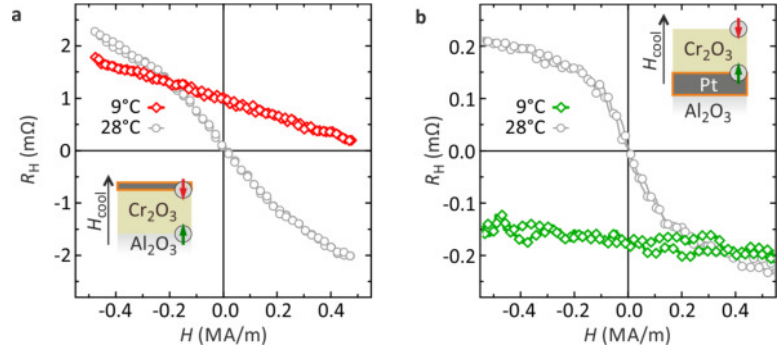

**Supplementary Figure 1 | Zero-offset Hall measurements on both surfaces of the $Cr_2O_3$ film. a,** Hall signal taken at the top surface becomes offset positively for cooling in a positive magnetic field. **b,** Hall signal taken at the bottom surface becomes offset negatively for cooling in a positive magnetic field.

These measurements confirm that the boundary magnetization at the $Cr_2O_3$(0001) top surface is approximately compensated by the antiparallelly aligned magnetization at the bottom surface. The ferromagnetic boundary layers thus do not conflict with perfect antiferromagnetism of $Cr_2O_3$ thin films or magnetoelectric antiferromagnet films in general. As a result, the areal magnetization at one boundary is expected to significantly larger than the integral ferrimagnetic moment density estimated by SQUID[16] or by the gate bias voltage [main text]. This further corroborates that the boundary magnetization is indeed largely compensated when taking the full thickness of the $Cr_2O_3$ film into account.

However, one can infer that the sublattice containing the bottom interface, has a larger magnetic moment, since the bottom interface magnetization aligns with the magnetic field, while the top interface is consistently aligned opposite to the magnetic field. As a result, the average AF order parameter of the $Cr_2O_3$ thin film becomes susceptible to medium strength magnetic fields of less than 1 MA/m.

### V. Isothermal magnetoelectric switching in the $V_2O_3$ gated sample

$V_2O_3$ offers interesting possibilities as a gate material for $Cr_2O_3$-based MERAM. It allows to avert twinning as shown in the main text, while providing a gate electrode. Its ultra-low lattice mismatch guarantees negligible ferrimagnetism in the magnetoelectric layer. **Supplementary Figure 2** shows a room temperature hysteresis loop of the AF order parameter in dependence of the gate voltage extending to 2 V. Despite the low applied gate voltage, the hysteresis opening is discernible. The shown curve is an average of 30 individual hysteresis loops, which leaves an uncertainty of about 5 µΩ in the individual data points. When comparing the upper and lower branches as wholes, the mean value of the upper branch is higher than that of the lower branch by $\Delta\langle R_H\rangle=(9.97\pm 0.53)$ µΩ. Therefore, the hysteresis loop is clearly open with a significance of $18.9\,\sigma$.

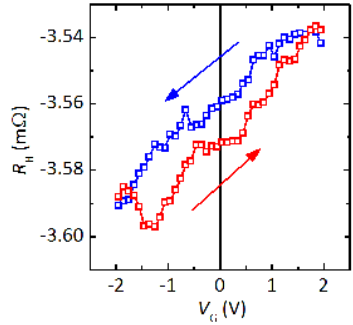

**Supplementary Figure 2 | Isothermal magnetoelectric switching of $Cr_2O_3$ on $V_2O_3$.** The hysteresis loop was obtained at 23°C while a permanent magnetic field of $H_{\mathrm{perm}} = 0.5$ MA/m was applied.

When comparing the behavior of the $V_2O_3$ gated cell in magnetoelectric field cooling [main text] and isothermal switching to similar tests carried out for the Pt gated cells in the main text, some conclusions can be drawn. The lower lattice mismatch provided by $V_2O_3$ reliably suppresses the gate bias. On the other hand, the magnetoelectric susceptibility of $Cr_2O_3$ appears to be reduced when grown on $V_2O_3$. The latter could be a result of doping due to cation exchange, which is a spontaneous process at high temperatures due to the excellent miscibility of the two materials. Optimized deposition conditions can be employed to alleviate this issue.

### VI. Structural investigation of the $Cr_2O_3$ films on different underlayers

The in-plane and out-of-plane crystalline order of epitaxial thin films can be efficiently probed by a 2-dimensional reciprocal space map, if one of the crystalline axes is selected as the in-plane direction. Through varying the incident and scattering angles, one obtains an overview such as that in **Supplementary Figure 3**. The maps are all aligned in the 3-dimensional reciprocal space by the 1 0 10 reflection of the $Al_2O_3$ substrate and thus one finds this reflection in all the maps. All the present reflections are identified in panel c. As $Al_2O_3$ is a single-crystalline substrate, its properties are highly consistent and the peak intensity is observed at the lattice position of relaxed $Al_2O_3$. The second reflection present in all the maps is that of $Cr_2O_3$. The appearance of reflections not belonging to the substrate is a clear indication that $Cr_2O_3$ is indeed crystalline and its position in reciprocal space clearly identifies it as a 1 0 10 reflection akin to the substrate reflection. To confirm the epitaxial relationship unambiguously, the in-plane symmetry can be assessed by probing the reciprocal space ring through the reflection and concentric about the [0 0 1] axis. The $Cr_2O_3$ and $V_2O_3$ films grown on $Al_2O_3$ substrates cannot maintain the threefold symmetry [**Supplementary Figure 3**(d-f)] entirely, which is caused by the formation of grains with a second in-plane orientation during film growth. The preponderance of these grains varies strongly between the different systems. We will refer to those grains as twin grains as they have the same crystallographic orientation as rotational twins. While all oxide systems contain only about 2% such grains, $Cr_2O_3$ films on Pt have no preference in the selection of the twin type.

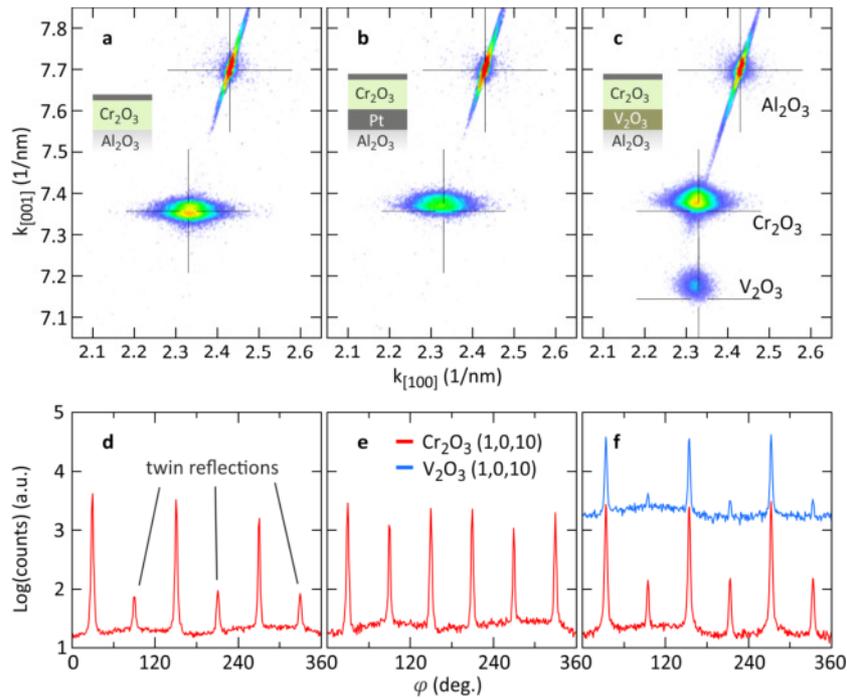

**Supplementary Figure 3 | Structure characterization by X-ray diffraction. a-c**, 2-dimensional reciprocal space maps aligned with the [1 0 0] substrate axis and spanning the range around the 1 0 10 reciprocal corundum lattice point. Black crosshairs indicate the relaxed room temperature lattice position of the three corundum oxides. **d-f**, Corresponding $\varphi$ scans through the $Cr_2O_3$ (red curve) and the $V_2O_3$ lattice points (blue curve), which reveal the present in-plane rotational symmetry.

The in-plane lattice misfit between $Cr_2O_3$ and $Al_2O_3$ is substantial at 4.0 %. It is remarkable that the position of the $Cr_2O_3$ reflection exactly matches its relaxed bulk value, which is overlaid as a black cross, when $Al_2O_3$ is used as a substrate. This implies that the misfit strain of $Cr_2O_3$ on $Al_2O_3$ is fully relaxed by misfit dislocations. As no appreciable signature of unrelaxed $Cr_2O_3$ is evident, the unrelaxed $Cr_2O_3$ volume must be negligible and the dislocations must be situated close to the $Al_2O_3/Cr_2O_3$ interface. The misfit leads to an average spacing of misfit lines of 25 $Cr_2O_3$ unit cells or 10.7 nm. At the same time, the $\varphi$ scan of the reciprocal lattice ring [**Supplementary Figure 3**(d)] shows that the 3-dimensional symmetry is almost perfectly transferred to the $Cr_2O_3$ layer, apart from a 2 % minority area where crystals with a 60° in-plane rotation are present. These twins give rise to the smaller intermediate reflections in the $\varphi$ scan. These data indicate that the $Cr_2O_3$ layer directly on $Al_2O_3$ obtains structural properties close to that of a single crystal although the lower interface region is substantially disturbed.

Similar measurements for the $Al_2O_3/Pt/Cr_2O_3$ system are shown in **Supplementary Figure 3**(b). In contrast to the $Al_2O_3/Cr_2O_3$ system, the $Cr_2O_3$ layers on the Pt buffer are found in a non-relaxed state after annealing and cooling down. Strain relaxation is a thermodynamically beneficial process which is limited by its slow kinetics at lower temperatures. Therefore, the maximally relaxed state is expected at the highest temperature experienced by the sample. The fully relaxed state observed in the $Al_2O_3/Cr_2O_3$ system implies that the annealing process is sufficient to obtain relaxed $Cr_2O_3$ at the annealing temperature.

The observed strain in the $Al_2O_3/Pt/Cr_2O_3$ system is developed during the cool-down to room temperature. Such thermal strains regularly arise when layers with different thermal expansion coefficients are stacked. The thermal expansion of $Al_2O_3$ between room temperature and 750°C is about 0.47 %[39]. That of $Cr_2O_3$ is expected to be similar, as $Al_2O_3/Cr_2O_3$ cools free of strain. Pt displays a larger thermal expansion of about 0.75 %[40]. When cooling the layer stack to room temperature, the expansion differences of the layers cannot be relaxed completely because the thermal energy of the cooling system is insufficient to invoke pronounced atomic rearrangements of the highly relaxed system. Instead, the layers retain an elastic strain if their thermal expansion ratios do not match. Due to the higher modulus and volume of the $Cr_2O_3$ layer compared to the Pt layer, the differential strain of the $Pt/Cr_2O_3$ system will be almost exclusively stored in the Pt lattice. Additionally, the complete $Pt/Cr_2O_3$ bilayer becomes tensely strained due to the differential $Al_2O_3/Pt$ strain of about 0.25 %. As a result, the $Cr_2O_3$ layer with the Pt seed layer retains an elastic c-axis compression of about 0.18 % at room temperature.

The presence of this elastic deformation indicates that the lattice misfit at the annealing temperature is a better indicator for the estimation of the misfit location density in the $Cr_2O_3$ films. This lattice mismatch between Pt and $Cr_2O_3$ is approximately 2.8 % giving rise to misfit lines every 14.9 nm. Thus, the Pt buffer reduces the linear density of misfit locations by about one third and the areal density by about one half. On the other hand, the $Cr_2O_3$ layer contains residual strain and displays severe twinning. In fact, the areal ratio between the two twins is about unity, implying equiprobable nucleation of each twin type.

$V_2O_3$ exhibits an even more pronounced thermal expansion between room temperature and 750°C of about 1.0 %[41]. While virtually perfectly lattice matched to $Cr_2O_3$ at room temperature, the high temperature misfit of $Cr_2O_3$ on $V_2O_3$ is about -0.5 %. When cooling to room temperature, the largest part of this differential strain of the $V_2O_3/Cr_2O_3$ is again stored in the underlayer, as the $Cr_2O_3$ film has a substantially larger volume. The differential $Al_2O_3/V_2O_3$ strain, however, affects both layers. The $V_2O_3$ lattice thus tries to contract about 0.5 % more than the $Al_2O_3$ lattice. As shown by the XRD data [**Supplementary Figure 3**(c)], only a small part of this thermal strain can still relax during the cool-down and the major part – about 0.3 % - is retained in the $Cr_2O_3$ lattice as an elastic $c$-axis compression. The strain stored in the $V_2O_3$ layer is expectedly even larger and amounts to 0.45 %.

Due to the subtle high temperature misfit, the $V_2O_3$ underlayer causes the lowest density of misfit dislocations in $Cr_2O_3$ with an average spacing of about 90 nm per misfit line. Compared to $Cr_2O_3$ on Pt, the linear misfit density is thus expected to be lower by a factor of about 6 using the $V_2O_3$ underlayer. The areal misfit density would be reduced by even the square of that factor. In further contrast to the Pt or $Al_2O_3$ underlayers, the misfit dislocations in the $V_2O_3$ system are expected to have an inverse stacking order, as the differential strain in the $Cr_2O_3$ layer is tensile at high temperatures with respect to the $V_2O_3$ lattice. At the same time, twinning is greatly suppressed when using a $V_2O_3$ underlayer as a gate electrode instead of a Pt layer [**Supplementary Figure 3**(f)]. The reason for this is that the corundum lattice is allowed to continuously extend from the substrate up to the $Cr_2O_3$ layer and no structure type boundary exists along the growth direction. Twinning mechanisms will be discussed in more detail below.

While the top surface of the magnetoelectric antiferromagnet is formed as a vacuum boundary during the annealing process, the situation for the bottom surface is more complex. The high temperature during annealing causes atom diffusion not only within $Cr_2O_3$, but potentially also an exchange of atoms between the underlayer and $Cr_2O_3$. Such interdiffusion occurs when $Cr_2O_3$ and the underlayer can form a solid solution or an alloy. Furthermore, during annealing at 750°C the $Cr_2O_3$ crystal relaxes any potential residual elastic strain apart from a small interfacial region, in which the interface bonds with the underlayer lattice disfavor the relaxed $Cr_2O_3$ lattice. As most of the $Cr_2O_3$ crystal tries to relax its stress, the misfit dislocations are transported to a region very close to its bottom interface, in case they formed further within the bulk.

When the underlayer is Pt, one potential effect is the dissolution of Cr into Pt. However, this process is strongly unfavorable in the case of $Cr_2O_3$ as, for Cr to dissolve, the oxide would need to be broken up and oxygen liberated. Therefore, a stoichiometrically sharp interface is expected between $Cr_2O_3$ and Pt[42]. In contrast, the $Al_2O_3$ or $V_2O_3$ underlayers are isostructural with $Cr_2O_3$ and are both oxides, which implies that $Cr_2O_3$ does not have to break up for intermixing to occur. Instead, the compounds mix by merely exchanging their trivalent metal cations. The miscibility depends on how similar the lattice constants are[43], implying more favorable and faster intermixing at $V_2O_3/Cr_2O_3$ interfaces. As the interfaces between the corundum oxides prepared at elevated temperatures can be considered a continued crystal with cation content gradually changing over some atomic distances[44,45], no atomically sharp interface exists. Therefore, the ferromagnetic boundary layer expected for free surfaces of $Cr_2O_3$ cannot arise in the same way at such interfaces.

While the particular effects differ between the various underlayers investigated here, the two unlike boundaries of the $Cr_2O_3$ layers introduce an inherent imbalance between the top and bottom ferromagnetic boundaries. As such magnetic surface effects are increasingly important when the film thickness is reduced, they must be considered when investigating the magnetic and magnetoelectric behavior of the thin $Cr_2O_3$ films. Eventually, ferrimagnetism as investigated throughout this manuscript can result from this magnetic moment imbalance.

The macroscopic degree of crystallographic twinning probed by XRD showed that the oxide thin films in all-corundum systems have a tendency to retain the 3-dimensional structural order of the $Al_2O_3$ substrate [**Supplementary Figure 3**]. The marked preference for one twin domain type over the other implies that the minority twins are stabilized merely by the presence of local defects that disturb the ideal $Al_2O_3$ surface. **Supplementary Figure 4** shows orientation contrast images to assess the microscopic pattern of the twin crystals. The contrast is obtained in an electron microscope by illuminating the surface along the oblique [1 0 4] direction. Under such a high symmetry direction, the atomic lattice contains channels that allow the probe electrons to enter far into the material before creating secondary electrons[46]. For the twin grain, these channels are rotated by 60° in the film plane and are thus not directly irradiated by the probe electrons. As a result, the twin type with more pronounced channeling will release less secondary electrons as they are created deeper inside the material.

**Supplementary Figure 4**(a) was obtained for an $Al_2O_3/V_2O_3/Cr_2O_3$ sample with only little twinning. The brighter minority twins are randomly distributed and their areal density is homogeneous at scales large than a few 100 nm. However, they vary substantially in size from below 10 nm up to 50 nm. As annealing the epitaxial film would rather remove small minority twin domains instead of creating them, the origin of the twinning must lie in the nucleation stage of film growth.

$Cr_2O_3$ films on non-corundum materials, on the contrary, reveal equiprobable twin domains as the seed layer exerts no preference for one particular epitaxial twin[27]. The twinning pattern then reveals the size of the individual islands that formed during the film growth. Due to the high mobility of Cr and O adatoms on (111) noble metal surfaces[42] the individual nucleations are substantially larger, than for $Cr_2O_3$ on other oxides, where bonding and sticking is stronger. As can be seen in **Supplementary Figure 4**(b) some $Cr_2O_3$ twin domains are over 100 nm in diameter. The image was taken for an $Al_2O_3$/Ni/Ag/$Cr_2O_3$ sample, which – like the Pt buffered films – contains twin domains of both types equiprobably.

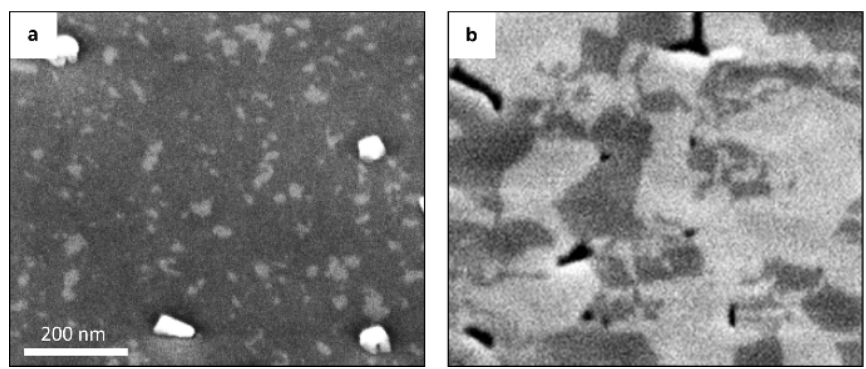

**Supplementary Figure 4** | $800 \times 800$ nm² **Electron channeling contrast micrographs** recorded using an electron illumination along the [1 0 4] direction of $Cr_2O_3$, which is tilted 31° off normal. **a,** $Cr_2O_3$ on a $V_2O_3$ underlayer shows an obvious preponderance of crystallographic twin domains with a dark contrast. **b,** $Cr_2O_3$ on a noble metal underlayer, in contrast, has a more equal occurrence of the two twin domain types.

Channeling contrast microscopy reveals that twinning is present in epitaxial $Cr_2O_3$(0001) films at scales of about 100 nm and below. On average, the observed twinning fraction found at such microscopic areas is consistent with the overall twin domain presence. Twinning does not affect the smoothness and continuity of epitaxial annealed $Cr_2O_3$ films. Therefore, twinned films are on equal footing with single orientation films from a structural point of view for the intents of this work.

As twinning evidently takes place on the scale of individual crystal nucleations regardless of the underlayer material, the approximate twin sizes are on the order of 100 nm. As the electrically probed regions are substantially larger with a diameter of about 10 μm, the twinning fraction within these regions is expected to be consistent with that measured by XRD for the entire sample.

**VII.  Theoretical background on using zero-offset Hall to probe ferrimagnetism**

The magnetic moment present in otherwise AF thin films of $Cr_2O_3$ can generate an order parameter selection pressure even when both the electric field and exchange bias are zero ($E = 0$, $J_{EB} = 0$). In this scenario, the film can be considered ferrimagnetic and the magnetic moment $m = \rho_m A$ contained in a particular area $A$ can invoke thermally activated selection of the magnetic order parameter. As the expected magnetic moment only arises as a small deviation from perfect AF order, the magnetic susceptibility of the order parameter is small and rather strong magnetic fields of several 100 kA/m are necessary for stable order parameter selection at room temperature. In the present thin film samples, a distribution of magnetic moments $\mathcal{M}(m)$ is expected due to both a natural spread of the size of the individual areas and a dependence of the areal magnetic moment density on the ordering temperature. The order parameter selection preference as a function of the applied magnetic cooling field is then described as a convolution:

$$\eta(H_{cool}) = \int_0^\infty \tanh\left(\frac{\mu_0 H_{cool} m}{k_B T_{crit}}\right) \mathcal{M}(m)\, dm \qquad (3)$$

The variation of the critical temperature among the individual domains can be neglected for the $k_B T_{crit}$ term as it has a spread of only about 1 % from its absolute mean value $\langle T_{crit}\rangle = 299$ K. The influence of this spread on the areal magnetic moment density is expected to be larger, but is already taken care of by the distribution $\mathcal{M}(m)$. Therefore, $T_{crit}$ will be regarded as a fixed parameter for this analysis. The distribution of the magnetic moments will be modeled by a Chi distribution

$$\mathcal{M}(m, \langle m\rangle, k) = X\left(\frac{m}{\langle m\rangle}, k\right) \qquad (4)$$

which describes a positive variable generated by an unknown number of degrees of freedom $k$. The Chi distribution is selected here, because it is a very general distribution with only two parameters and has several relevant special cases such as the Normal, Rayleigh or Maxwell distributions.

Using Eq. (3), it is possible to construct fits to experimental data of the field-invariant zero-offset Hall signal in dependence of the magnetic cooling field $R_{inv}(H_{cool})$. This has been done for the three main categories of samples [**Table 2**] and for one system without a Hall cross on top of the $Cr_2O_3$ layer. The latter was investigated by using the unpatterned Pt(20 nm) underlayer as a proximity magnet, which is possible due to the robust rejection of geometric Hall cross asymmetries of the zero-offset Hall technique[11]. The fits depend on three parameters, namely the two parameters $\langle m \rangle$ and $k$ describing the Chi distribution of the magnetic moments and the saturation Hall resistance $R_N$.

The cooling field dependences [left column in **Supplementary Figure 5**] of all the studied systems are well accounted for by the developed model [Eq. (3)], which yields a representation of the data that is essentially free of systematic errors. While the top-measured $Al_2O_3/Cr_2O_3$/Pt and Pt/$Cr_2O_3$/Pt systems reveal a marked positive dependence – and thus a positive magnetic moment – the bottom-measured Pt/$Cr_2O_3$/air and the top-measured $V_2O_3/Cr_2O_3$/Pt systems show inverse cooling field dependences of the zero-offset Hall signal. This inversion is caused by a net negative moment with respect to the sign of the magnetization in the probed boundary layer. In this respect, it is noteworthy that the dependences measured at the top and bottom boundaries in the Pt/$Cr_2O_3$/Pt system, respectively, are highly similar [**Supplementary Figure 5**(d,f)] apart from the inversion. Indeed, the two systems are expected to be structurally very similar, as the only difference is the deposition of the top Pt layer at room temperatures after the annealing of the magnetoelectric antiferromagnet layer. Both boundaries of the $Cr_2O_3$ layers in the top-measured Pt/$Cr_2O_3$/Pt and the bottom-measured Pt/$Cr_2O_3$/air system are thus prepared under identical conditions. The sign inversion of the field-invariant zero-offset Hall signal is thus a clear fingerprint of the fact that different sublattices of the antiferromagnet dominate the top and the bottom boundary of the magnetoelectric antiferromagnet layer, which is indeed in line with theoretical predictions[17]. The striking similarity in the magnetic field behavior of the order parameter selection [**Supplementary Figure 5**(d,f)] also implies that the room temperature deposition of the thin top Pt layer after the annealing process of the $Cr_2O_3$ layers has a negligible influence on the magnetic properties of the top boundary. Furthermore, this result also indicates that the magnetic moment of the bottom boundary sublattice is larger because the bottom layer is observed to align along the magnetic cooling field, while the top layer is aligned antiparallel to the cooling field direction. This conclusion is based on the observedly negative anomalous Hall signal of the $Cr_2O_3$/Pt proximity system [**Supplementary Figure 1**].

Before discussing the influence of the three different underlayers and hence the three sample categories [**Table 2**], it is important to consider the meaning of the individual fitting parameters listed in the right column panels of **Supplementary Figure 5**. The saturation value of the field-invariant zero-offset Hall signal $R_N$ is affected by the magnetic moment in the magnetic boundary and by the quality and thickness of the proximity magnet. In the present systems, the magnetic moment of the magnetoelectric AF boundary layer can be assumed to be similar for the bottom and top termination, unbalanced only to a few percent. As all the magnetic layers measured by Hall are epitaxial $Cr_2O_3$/Pt interfaces, the structural configuration can be also assumed to be similar. The remaining influence on $R_N$ is thus mainly related to the thickness induced signal shunting behavior of the Pt layer. This explains the about one order of magnitude lower value of $R_N$ for the Pt/$Cr_2O_3$/air system in comparison to the other systems, because the Pt underlayer has a thickness of about 20 nm compared to about 2.5 nm for the Pt top layers used in the other systems. In contrast to $R_N$, which is determined by the properties of the interface probed by Hall, the parameters $\langle m \rangle$ and $k$ describe the entire magnetic moment of the $Cr_2O_3$ layer over a particular area that selects an AF order parameter. The sign of $\langle m \rangle$ indicates if the total magnetic moment has the same or antiparallel orientation as the proximity magnetization. While all three parameters ($R_N$, $\langle m \rangle$ and $k$) are usually well confined [**Supplementary Figure 5**(b,d,f)], the measured dependence for the $V_2O_3$/$Cr_2O_3$/Pt system [**Supplementary Figure 5**(g)], is essentially linear with a rather low amplitude. This renders $R_N$ and $\langle m \rangle$ redundant parameters and makes $k$ badly determined. To obtain a meaningful interpretation, it is thus necessary to include the electric field as another stimulus [main text] and thus fix the saturation value $R_N$.

The general appearance of magnetic field susceptibility of the AF order parameter in all the studied magnetoelectric antiferromagnet thin film systems unambiguously demonstrates that the $Cr_2O_3$ films are not perfectly AF but have two slightly unbalanced sublattices giving rise to ferrimagnetism, even with no applied electric field. As there is no intrinsic cause for the broken sublattice equivalence, the parasitic magnetic moment must arise as a result of the extrinsic sample properties. Through careful review of the engineered differences between the three sample categories, it is possible to draw conclusions on which extrinsic properties drive the appearance the ferrimagnetism in magnetoelectric antiferromagnet thin films.

By measuring the zero field point in the dependences in **Supplementary Figure 5** many times, it is possible to determine the scatter induced due to the finite number of selectable domains within the Hall cross. This is shown in **Supplementary Figure 6**. The shown kernel density distributions are already deconvolved to remove the influence of the measurement accuracy. The remaining spread is purely a consequence of the domain discretization within the Hall cross and can be used directly to judge the relative domain sizes within the $Cr_2O_3$ layers on the various underlayers. Therefore, it is possible to measure both the domain moment, and the domain sizes using zero-offset Hall, which allows one to derive the ferrimagnetic areal moment density in the films. For this manuscript, only relative domain sizes will be used as the conclusion does not depend on the absolute sizes.

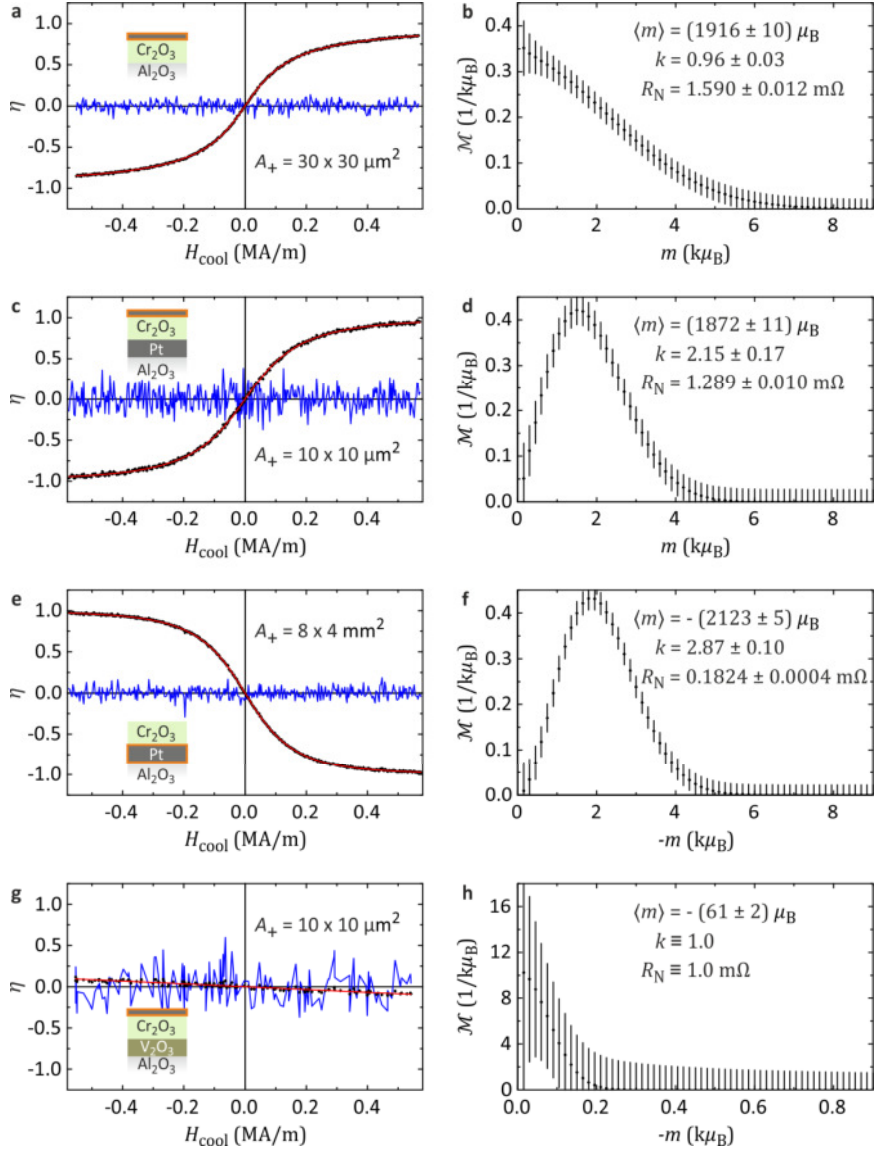

**Supplementary Figure 5 | a,c,e,g**, Experimental dependences of the AF order parameter on the magnetic cooling field for four different samples with different bounding layers as shown in the insets. Zero-offset Hall measurements have been carried out on the orange highlighted layer. Red lines show fits obtained using Eq. (3) and blue lines are residuals x10. **b,d,f,h**, Corresponding magnetic moment distributions as obtained from the fits.

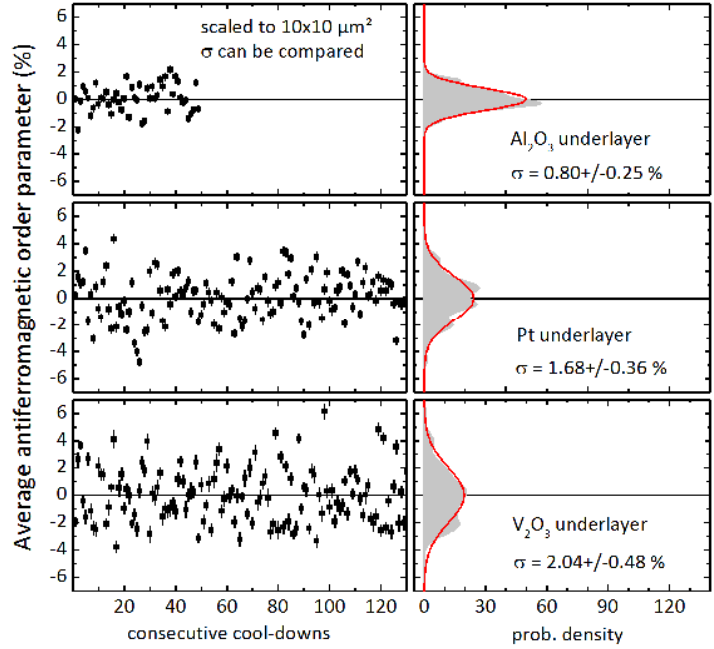

**Supplementary Figure 6 | Relative domain size measurement.** By studying the variations between identical cool-downs, one can conclude on the relative average domain sizes.

## VIII. Positron Annihilation Spectroscopy

The fate of positrons in solids is to thermalize, diffuse, and then annihilate with core and valence electrons of the material, which results in emission of two ~511 keV gamma photons. Due to the momentum of the electrons, the variation from that value is a result of Doppler broadening (DB) of the annihilation line. The Doppler broadening is characterized by the shape parameter $S$ and the wing parameter $W$. More details about both parameters can be found elsewhere[47,48], but in general the profile of these parameters in terms of positron implantation depth is influenced by the stopping profile $P(z, \epsilon)$ and positron diffusion. The $S$ parameter is more sensitive to the open volume defects concentration and their size, whereas the $W$ parameter is a fingerprint of the annihilation site surrounding.

Here, we employ a DB positron annihilation spectroscopy (PAS) setup for defect concentration depth profiling of a 170 nm thick epitaxial $Cr_2O_3$ film on an $Al_2O_3$ substrate (thickness determined by cross-sectional TEM). The depth sensitivity is given by the variation of positron incident energies. The positron stopping profiles are approximated by a Makhovian distribution and the mean positron penetration depth is $z_{mean} = A \rho^{-1} \epsilon^n$, where $\epsilon$ is the positron energy, $\rho$ is the material density, and the parameters $A$ and $n$ are material-related constants. $z_{mean}$ has been calculated for $Cr_2O_3$ and can be found in **Supplementary Figure 7**(a) as the top scale. The maximum positron implantation depth is about $2 \cdot z_{mean}$. The $S(\epsilon)$ curve indicates an increased open volume at an implantation energy of 7 keV, which corresponds to $z_{mean} \approx 162$ nm, thus very close to the expected interface position. The black line at $\epsilon = 5$ keV bounds the regime, in which basically all positrons annihilate in the $Cr_2O_3$ film only. The $S(\epsilon)$ dependence has been fitted with the VEPFIT code[49] in order to calculate the effective diffusion lengths in the system as well as the characteristic $S$-parameters. The fit shows a good agreement with experimental data.

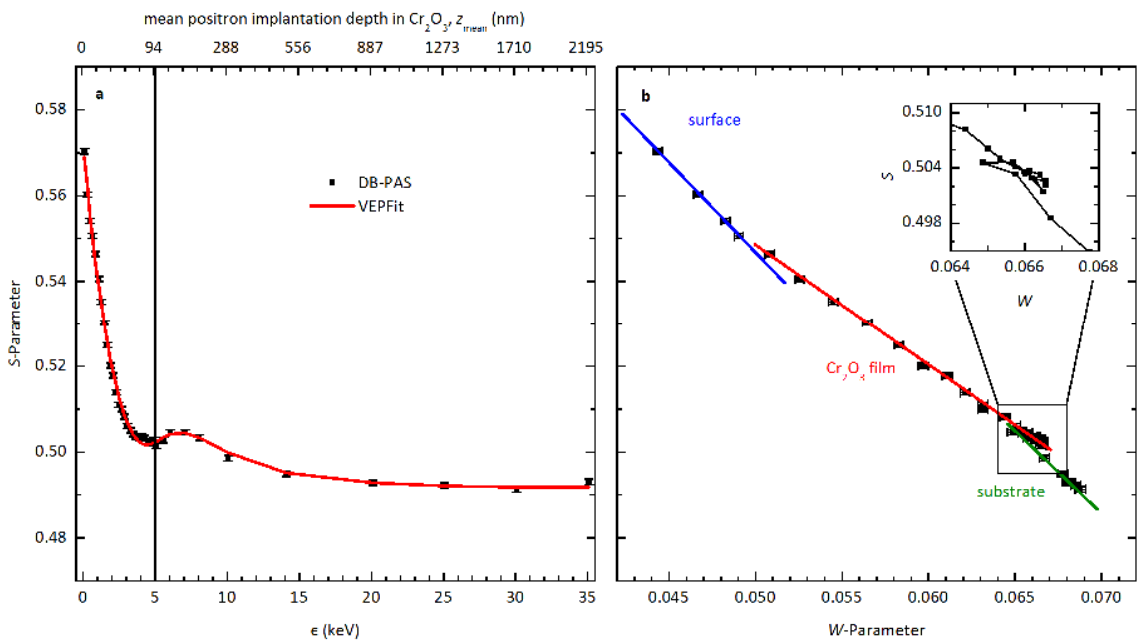

**Supplementary Figure 7 | Positron Annihilation Spectroscopy of $Cr_2O_3$ thin films.** a, Annihilation line parameter S, as a function of positron energy $\epsilon$ (black squares) with corresponding positron diffusion modeling (red line) extracted from the VEPFIT code. b, S-W plot clearly showing three distinct regions across sample thickness, namely a thin surface region, bulk, and a substrate. In the inset, the interface between the film and substrate is shown in more detail – the characteristic zig-zag shape of the curve indicates increased open volume at the interface.

Model 5 of VEPFIT is used to fit the data, where a rectangular distribution of defects and a layered structure are assumed. Four distinct regions have been considered, three of them with thicknesses of 10 nm, 159 nm, and 1 nm for a film-surface region, film-bulk, and film-interface, respectively. The fourth region is the substrate. The film-surface region has been recognized from the $S$-$W$ plot [**Supplementary Figure 7**(b)], where the four data points corresponding to the lowest implantation energy lie on a slope different than that of the data taken at higher energies. This suggests a different defect type in the topmost 10 nm of the film, according to the Makhovian distribution. The other distinct regions are well discernible in the $S$-$W$ plot, too. The interface between the $Cr_2O_3$ film and the substrate is especially characteristic [inset of **Supplementary Figure 7**(b)], with a zig-zag shape of the $S$-$W$ curve clearly visible. All the characteristic regions of the $S$-$W$ plot can be fitted with lines of different slopes, which translates to different defect types. The fitting zone for the VEPFIT calculation is divided into 49 depth intervals starting with 0.1 nm and with an increment factor of 1.3. The best fit values for the $S$ parameters are summarized in **Supplementary Table 1**. The epithermal scattering length is given at (2.5 ± 0.08) nm. The effective positron diffusion lengths for each of the four regimes have been fixed in the calculation yielding the best fit for values of about 5 nm, 20 nm, 1 nm, and 60 nm for the film-surface-region, film-bulk, film-interface, and the substrate, respectively. The diffusion length for the substrate is in agreement with previous reports[50]. The calculated $S$-parameters of the layers reveal a significantly enhanced trapping potential at the film-interface region, corresponding to an enhanced defect concentration at the $Cr_2O_3/Al_2O_3$ interface than in the bulk of the $Cr_2O_3$ film. The positron trapping in the film is likely due to mono-vacancies that are difficult to detect with other measurement techniques.

|  | surface | film-surface | film-bulk | film-interface | substrate |
|---|---|---|---|---|---|
| $S$-parameter | 0.5452 (±.0020) | 0.5405 (±0.0014) | 0.49359 (±0.00031) | 0.5378 (±0.0018) | 0.49150 (±0.00029) |
| diffusion length (nm) |  | 5 | 20 | 1 | 60 |

**Supplementary Table 1 | $S$-parameters and diffusion lengths of distinct layers in the $Cr_2O_3/Al_2O_3$ system calculated by VEPFIT.**